\documentstyle[prl,aps,twocolumn]{revtex}

\begin{document}
\draft
\title{Magnetic Field Scaling of the Conductance of Epitaxial Cuprate-Manganite
Bilayers.}
\author{P. A. Kraus*$^{\dagger }$, A. Bhattacharya$^{\dagger }$, and A. M. Goldman}
\address{School of Physics and Astronomy, University of Minnesota, \\
116 Church St. SE, Minneapolis,\\
MN 55455, USA}
\date{July 20, 2001.}
\maketitle

\begin{abstract}
Conductance-voltage characteristics of epitaxial interfaces between oxide
ferromagnets and oxide superconductors have been measured as a function of
temperature and magnetic field. Their functional form is similar to that
predicted by theories of transport across nearly transparent contacts
between highly spin-polarized ferromagnets and {\it d-}wave superconductors.
However, their magnetic field dependencies scale in striking and unusual
ways, challenging our current understanding. Existing theories fail to
account for apparent nonequilibium effects that are natural for spin
injection in such geometries.
\end{abstract}


The half-metallicity of the alkaline earth doped lanthanum manganites leads
to their charge carriers being spin-polarized \cite{ramirez}. This, together
with the epitaxial compatibility of these oxide ferromagnets and oxide
superconductors \cite{chahara}, has led to the use of such heterostructures
for the study of spin injection \cite{johnson} in the cuprates \cite
{vaskoprl,vaskoapl,marylandapl,stroud,yehprb}. This has motivated a number
of theoretical papers \cite{zvalls,kash,ting,maekawa,golubov,klapwijk}, most
of which describe interfaces of varying transparency exhibiting suppression
of Andreev reflection, but except for Refs.\cite{golubov} and \cite{klapwijk}
do not consider nonequilibrium effects. In this Letter we report on
conductance-voltage characteristics, {\it G(V)}, of cuprate/manganite
heterostructures, with no buffer layer between the cuprate and manganite.
The geometry is arranged so that the current density is uniform across the
area of contact between the superconductor and the ferromagnet. The
zero-bias conductance peak (ZBCP) reported by other groups \cite
{greenezbcp,sawazbcp,chenzbcp} is absent. Furthermore, as a function of
temperature and magnetic field, striking and unusual scaling of the data at
low bias is found. This observation puts a serious constraint on any theory,
and conversely, a theory that explains all the features of the data might
allow for the intriguing possibility of gap spectroscopy of superconductors
using such structures. A complete description would have to account for
nonequilibrium effects of spin injection on the underlying superconductor.

Epitaxial heterostructures consisting of c-axis oriented DyBa$_{2}$Cu$_{3}$O$%
_{7-\delta }$ (DBCO, 50 nm thick) and La$_{2/3}$Ba$_{1/3}$MnO$_{3}$ (LBMO,
40 nm thick) capped with Au (50 nm thick) were grown {\it in situ} on
epitaxially polished SrTiO$_{3}$ substrates using ozone-assisted molecular
beam epitaxy in the block-by-block mode \cite{locquet,growth}. The as-grown
buried DBCO layer was not superconducting, necessitating an oxygen annealing
step, which took a few days. The superconducting transition temperatures ($%
T_{c}$) of the DBCO layers were all about 70 K, a consequence of their being
underdoped. The interfaces were of high quality, with cross-sectional
transmission electron microscopy demonstrating heteroepitaxy of the layers
over semi-macroscopic distances, and {\it in situ} RHEED studies carried out
during growth suggesting very abrupt transitions between layers.

The Au and LBMO were patterned into rectangles 10 $\mu $m wide, with varying
lengths, on a 10 $\mu $m wide strip of the underlying DBCO film (Fig.1).
Electrical leads were patterned for vertical four-terminal measurements of $%
G(V)$ through each structure. The Au layer provided an equipotential
surface, and forced the current to traverse the numerous interfaces in a
vertical direction (parallel to the {\it c}-axis). During such measurements,
no current flowed in the plane of the manganite layer. Additional leads
permitted simultaneous four-terminal measurements of the superconducting
layer.

Figure 2 shows plots of {\it G(V)}, at several temperatures of a structure,
10 $\mu $m $\times $ 20 $\mu $m in area. Not shown is the
voltage-independent conductance observed above $T_{c}$. There are several
features of these curves, which should be noted: below $T_{c}$ a zero-bias
conductance peak appears. As the temperature is reduced, this transforms
into a plateau, and then into a ''vee''-shaped dip, which deepens with
decreasing temperature. The maxima in {\it G(V)} at $\pm $40 mV found at low
temperatures are followed by a rapid fall off, while for a 10 $\mu $m $%
\times $ 10 $\mu $m structure, they occurs at about $\pm $80 mV. These
features cannot be associated with the superconducting gap amplitude as
their value would be unphysically large for an underdoped film. In fact, the
observed maxima actually mark the beginning of the demise of
superconductivity due to current injection. At 2 K, the underlying DBCO
strip is driven normal by an injection current through the ferromagnet of
about 66 mA, while 86 mA is required if injection is in the plane. Control
experiments in which the manganite was replaced by either Au or LaNiO$_{3}$
films in similar geometries and with the same resistances show no nonlinear
effects in {\it G(V)} at similar current densities. This supports the view
that the effects are not a consequence of heating and are due to spin rather
than charge injection \cite{vaskoapl,marylandapl}. Estimates of power
dissipation under injection conditions suggest a temperature rise of at most
1 K, which is also not sufficient to account for the observations. The inset
shows a monotonous increase in the zero-bias resistance of the trilayer on
cooling below $T_{c},$ contrary to what one would expect for resistance at
grain boundaries in the manganite\cite{todd}.

The magnetic field dependences of {\it G(V)}, at temperatures from 2 K to 70
K are shown in Fig.3. The field is applied in the plane of the
cuprate/manganite interface. The qualitative effect of increasing field is
similar to that of increasing temperature. This is dramatically borne out by
the inset. Similar results were reported in our earlier work on geometries
in which the current density was not uniform\cite{vaskoapl}. There is a
small asymmetry in the {\it G(V)} which grows with increasing magnetic field
and decreasing temperature. [{\it G(V+) - G(V-)]/G(V)} is less than 5\% at
12 T and 2 K.

Notable by its absence at low temperatures is the zero-bias conductance peak
(ZBCP) that has been reported in other experiments \cite{sawazbcp,chenzbcp}.
This peak is usually attributed to sub-gap Andreev bound states, which occur
for tunneling in the {\it a-b} directions (maximally along (110) directions)
into a cuprate superconductor, because of the change of sign in the {\it d-}
wave order parameter on rotation by an angle of $\pi {\it /}2$ in the {\it %
a-b} plane \cite{greenezbcp}. The present epitaxial {\it c}-axis oriented
films are relatively free of such {\it a-b} facets at the cuprate/manganite
interface. For spin polarized injection in this geometry, the effective gap
is an angular average over all directions, and this suppresses the ZBCP.
Also, for an interface with high transparency, one would expect the ZBCP to
disappear.

A very striking and unusual feature of the data is shown in Fig. 4, where we
have plotted {\it G(V,H)/G(0,H)} for different values of $H$ at a fixed
temperature. All curves in various magnetic fields collapse onto a single
curve at a fixed temperature. This works at each temperature, but with a
functional form that is different in detail. The difference may be due to an
increase of the quasiparticle population in the superconductor with
temperature, and/or a change in the spin polarization of carriers in the
manganite\cite{spinpol}. This collapse at low bias has been observed in all
devices measured, at all fields and temperatures where a minimum in $G(V)$
is clearly seen. This implies that at low bias voltages, {\it %
G(V,H,T)=h(H,T)g(V,T)}. At a fixed temperature, the collapsed curve is
essentially {\it g(V,T)}. The field dependent prefactor, normalized by its
value at zero field, i.e. {\it h(H,T)/h(0,T)}, is shown in Fig. 5. At the
lowest temperatures of 2 K and 10 K, it is fit very well by a straight line.
In fact, the 2 K and 10 K data lie nearly on top of each other. At higher
temperatures, though still very close to linear,{\it \ h(H,T)} deviates
slightly from the straight line fit at 2K. This kind of collapse is also
suggested by the data of Sawa {\it et al}. at high bias voltages\cite
{sawazbcp}, although at low bias it is obscured by the ZBCP. In our data we
also observe that plots $G(0,T)/G(0,T=2K)$ at all applied magnetic fields,
seem to have very similar temperature dependences (see Fig. 5, inset).

We note that the zero-bias conductance $G(0)$ increases by about a factor of
two when temperature is increased from 2 K to 10 K, keeping the magnetic
field fixed. Correspondingly, on increasing the magnetic field up to 12 T
and keeping the temperature fixed, the zero-bias conductance again increases
by a factor of about two. Considering that ${\it \mu }_{B}\Delta {\it H}%
\approx {\it \ k}_{B}\Delta {\it T}$ for $\Delta {\it H}=$12 T and $\Delta 
{\it T}=$ 8 K, one might conclude that the measurements are probing the
density of states near the nodes \cite{kash} of a {\it d-}wave like gap,
with the increase in conductance corresponding to an increase in the number
of accessible states. If there is suppression of Andreev reflection at the
ferromagnet/superconductor interface, then {\it G(V) }probes the density of
states of the superconductor\cite{kash}. Motivated by this, and presuming a
small barrier at the cuprate/manganite interface, we have calculated
temperature and magnetic field dependence for the conductance-voltage curves
at the interface between a half-metallic ferromagnet and a {\it d-}wave
superconductor using a simple tunneling expression\cite{meservey}

\begin{equation}
I(V)\sim \int_{-\infty }^{\infty }\nu _{fm}\nu _{s}(E+\mu
H)\{f(E+eV)-f(E)\}dE.  \label{one}
\end{equation}
Here the density of states for quasiparticles in a {\it d-}wave
superconductor $\nu _{s}(E)\sim \nu _{n}(E$/$\Delta _{o})$, $\Delta _{o}$ is
the gap amplitude, $f$ is the Fermi function, and $\nu _{n}$ is the density
of states at the Fermi energy of the superconductor in the normal state. The
density of states in the ferromagnet $\nu _{fm}(E)$ is assumed constant.
Applying a magnetic field enhances the density of states for quasiparticles
of one spin species at the expense of the other, enhancing conductance for
carriers of the predominant spin across the interface. We plot $G(V)=$ $%
dI/dV $ as a function of applied magnetic field and temperature at low
applied bias (Fig. 6). {\it G(V)} increases linearly as a function of {\it H}
at all bias voltages. This suggests that the spin polarization of carriers
plays a role in the enhancement of conductance in the experiment. The
calculated increase in $G(0)$ is the same when {\it T} increases from 2 K to
10 K with {\it H} = 0, and when {\it H} increases from 0 T to 12 T at {\it T}
= 2 K, which is by a factor of about five. This increase is greater, though
of the same order of magnitude as that seen in the experiment. However, the
scaling of the experimental data indicates that the application of magnetic
field has a {\it multiplicative} effect on the conductance, with the
increase being greater at higher bias voltages. This leads us to believe
that the explanation of this phenomenon involves physics different from that
of our simple argument.

In choosing to ensure that currents flowed vertically through the structure,
we added a gold overlayer. This compromised our ability to make true four
terminal measurements of the cuprate/manganite interface, as the resultant
gold/manganite series resistance cannot be subtracted in any systematic
fashion. The 10 $\mu $m $\times $ 20 $\mu $m gold/manganite/cuprate trilayer
structure has a zero-bias resistance of about 1.46 $\Omega $ at 10 K. From
other measurements, the gold/manganite interface resistance is known to be
ohmic. Assuming that the linear field dependence of {\it G(V,H) }occurs due
to the cuprate/manganite interface, the field independent ohmic part of the
trilayer resistance can be extracted from the data near zero-bias, and is
about 0.06 $\Omega $ at 10 K. The gold/manganite interface resistance is
presumably a fraction of this value. At higher injection currents where the
trilayer resistance ({\it V/I}) is about 0.5 $\Omega ,$ the contact
resistance of the gold/manganite interface is a greater fraction of the
total resistance, and the actual voltage across the cuprate/manganite
interface near the peaks in {\it G(V) }may be significantly smaller than the
measured voltage.

Given the values of specific interface resistances (gold/manganite +
manganite/cuprate), which were of order $10^{-6}$ $\Omega -cm^{2}$, the
interfaces are highly transparent, but certainly not ballistic. Thus the
striking voltage dependence of {\it G(V)} at low bias, although reminiscent
of suppressed Andreev reflection\cite{zvalls,kash,ting}, cannot be explained
by that alone. Furthermore, injection of spin-polarized quasiparticles into
a superconductor creates a steady-state spin imbalance analogous to the spin
imbalance created by the application of a magnetic field due to Zeeman
splitting of the spin energies\cite{maekawa}. This leads to a suppression of
the superconducting order parameter near the manganite/cuprate interface.
Also, the voltage drop due to injection of these spin-polarized
quasiparticles would then occur over a characteristic relaxation length in
the superconductor, and not abruptly at the interface (as in the case of
tunneling). If this relaxation length is sufficiently large in the cuprate,
the measured voltage would not be the spectroscopic voltage of the
interface. Thus, the absence of any signatures of the coherence peak in {\it %
G(V)} may be because the superconductor near the boundary is driven normal
before the voltage across the manganite/cuprate interface reaches a value
corresponding to the gap amplitude. This situation is not part of any of the
theories as the pair potential is assumed to be constant as a function of
distance from the interface, and independent of current.

This work was supported by the Office of Naval Research under Grant
N/N00014-98-1-0098. The authors would like to thank J.P. Locquet,V.
Vas'ko,V. Larkin, A.I. Larkin and L.I. Glazman for useful discussions.

\begin{description}
\item  {\footnotesize *Present address: Applied Materials Inc., Santa Clara,
CA. }$^{\dagger }${\footnotesize Authors contributed equally to this work.}
\end{description}

\begin{figure}[tbp]
\caption{Device geometry for vertical spin injection. I2/V2 and I3/V3 are
used to measure the LBMO/DBCO interface, while I1/V1 and I3/V3 are used to
characterize the DBCO.}
\label{Fig.1}
\end{figure}

\begin{figure}[tbp]
\caption{$G(V)$ for a Au/LBMO/DBCO heterostructure, of area $10\mu m$ $%
\times $ $20\mu m$. The curves show a conductance peak at the highest
temperature. At low temperature, the $G(V)$ data are rounded at low bias,
tending to two arms of a `vee' shape at higher bias, before falling abruptly
as the underlying superconductor is quenched. The inset shows the zero bias
conductance of the trilayer as a function of temperature.}
\label{Fig.2}
\end{figure}

\begin{figure}[tbp]
\caption{$G(V)$ in magnetic fields applied parallel to the plane for a 10 $%
\mu $m $\times $ 20 $\mu $m device at 10 K. The inset shows data at 65 K.}
\label{Fig.3}
\end{figure}

\begin{figure}[tbp]
\caption{Conductance data at 10 K normalized by the zero bias conductance at
each magnetic field, plotted vs.voltage for the 10 $\mu $m $\times $ 20 $\mu 
$m device. The data collapse onto a single curve at low bias for all applied
magnetic fields up to 12 T. }
\label{Fig.4}
\end{figure}

\begin{figure}[tbp]
\caption{$G(0)$ for the 10 $\mu $m $\times $ 20 $\mu $m device plotted vs.
magnetic field H applied parallel to the plane, at different temperatures.
The data are normalized to the conductance at H = 0. A straight line is fit
to the data at $2K$ as a guide to the eye. The inset shows $G(0)$ plotted as
a function of temperature at different values of magnetic field. The data at
each magnetic field are normalized to the value of the conductance at T = 2
K.}
\label{Fig.5}
\end{figure}

\begin{figure}[tbp]
\caption{$G(V)$ calculated using density of states form from Eq. 1. for spin
injection from a half-metallic ferromagnet. The change in conductance on
applying magnetic field from 0 T to 12 T at 2 K is compared to the change
caused by increasing the temperature from 2 K to 10 K at 0 T(inset).}
\label{Fig.6}
\end{figure}




\begin{references}
\bibitem{ramirez}  For a review see: A. Ramirez, J. Phys.: Condens. Matter 
{\bf 9}, 8171 (1997).

\bibitem{chahara}  K. Chahara {\it et al.}, Appl. Phys. Lett. {\bf 63}, 1990
(1993).

\bibitem{johnson}  Mark Johnson and R. H. Silsbee, Phys. Rev. B {\bf 37},
5326 (1988); The first spin injection study of a cuprate involved the used
of Permalloy. See: Nir Hass {\it et al}., Physica C {\bf 235-240}, 1905
(1994).

\bibitem{vaskoprl}  V. A. Vas'ko {\it et al}., Phys. Rev. Lett. {\bf 78},
1134 (1997).

\bibitem{vaskoapl}  V. A. Vas'ko {\it et al}., Appl. Phys. Lett. {\bf 73},
844 (1998).

\bibitem{marylandapl}  Z. W. Dong {\it et al}., Appl. Phys. Lett.{\bf \ 71},
1718 (1997).

\bibitem{todd}  N.K. Todd {\it et al., }J. App. Phys. {\bf 85}, 7263 (1999).

\bibitem{stroud}  R. M. Stroud {\it et al}., J. Appl. Phys. {\bf 83}, 7189
(1998).

\bibitem{yehprb}  N-C. Yeh {\it et al.}, Phys.Rev. B {\bf 60}, 10522 (1999).

\bibitem{zvalls}  I. Zutic and O. T. Valls, Phys. Rev. B {\bf 60}, 6320
(1999); I. Zutic and O. T. Valls, Phys. Rev. B {\bf 61}, 1555 (2000).

\bibitem{kash}  S. Kashiwaya {\it et al.}, Phys. Rev. B {\bf 60}, 3572
(1999); Y. Tanaka and S. Kashiwaya, Phys. Rev. Lett. {\bf 74}, 3451 (1995);
J. H. Xu, J. H. Miller and C. S. Ting, Phys. Rev. B.{\bf \ 53}, 3604 (1996).

\bibitem{ting}  Jian-Xin Zhu, B. Friedman, and C. S. Ting, Phys. Rev. B. 
{\bf 59}, 9558 (1999).

\bibitem{maekawa}  S. Takahashi, H. Imamura, and S. Maekawa, Phys. Rev.
Lett. {\bf 82}, 3911 (1999).

\bibitem{golubov}  A.A. Golubov, Physica C {\bf 326-327}, 46 (1999).

\bibitem{klapwijk}  F.J. Jedema {\it et al.}, Phys. Rev. B{\bf \ 60}, 16549
(1999).

\bibitem{meservey}  R. Meservey and P.M. Tedrow, Phys. Rep. {\bf 238, }%
173-243 (1994).

\bibitem{greenezbcp}  M. Covington {\it et al.}, Phys. Rev. Lett. {\bf 79},
277 (1997).

\bibitem{sawazbcp}  A. Sawa {\it et al.}, cond-mat/9908431.

\bibitem{chenzbcp}  Z. Y. Chen {\it et al}., cond-mat/0007353.

\bibitem{locquet}  J.-P. Locquet {\it et al.,} Appl. Phys. Lett. {\bf 64},
372 (1994).

\bibitem{growth}  V. A. Vas'ko {\it et al.}, Appl. Phys. Lett. {\bf 68},
2571 (1996).

\bibitem{spinpol}  J. Y. T. Wei, N.-C. Yeh, and R. P. Vasquez, Phys. Rev.
Lett. {\bf 79}, 5150 (1997); J-H. Park {\it et al}., J. Appl. Phys. {\bf 79}%
, 4558 (1996); Moon-Ho Jo {\it et al.}, Phys. Rev. B {\bf 61}, 1495 (2000).
\end{references}
\end{document}